\documentclass[prd,aps,preprint,epsfig,floats]{revtex4}
\usepackage{graphics}
\usepackage{epsfig}
\newcommand{\be}{\begin{equation}}
\newcommand{\ee}{\end{equation}}
\newcommand{\bea}{\begin{eqnarray}}
\newcommand{\beas}{\begin{eqnarray*}}
\newcommand{\eea}{\end{eqnarray}}
\newcommand{\eeas}{\end{eqnarray*}}
\newcommand{\ba}{\begin{array}}
\newcommand{\ea}{\end{array}}
\newlength{\myem}
\newcounter{mysubequation}[equation]

\newlength{\phantomlength}

\makeatletter
\@addtoreset{equation}{section}
\makeatother

\begin{document}

\draft
%@twocolumnfalse\endcsname
\preprint{\vbox{\hbox{UMD-PP-07-003}}}

\bigskip
\bigskip

%
%\title{ Natural Realizations of Seesaw in Mini-Warped Minimal SO(10) }
%\author{R. N. Mohapatra}
%\email{rmohapat@physics.umd.edu}\affiliation{Department of Physics,
%University of Maryland, College Park, MD, 20742}
%\author{Nobuchika Okada}
%\email{okadan@kek.jp} \affiliation{KEK, Tsukuba, Japan}
%\author{Hai-Bo Yu}
%\email{hbyu@physics.umd.edu}\affiliation{Department of Physics,
%University of Maryland, College Park, MD, 20742}
%

\title{ Natural Realizations of Seesaw in Mini-Warped Minimal SO(10) }

\author{R. N. Mohapatra\footnote{
  e-mail: rmohapat@physics.umd.edu},
        Nobuchika Okada\footnote{
  e-mail: okadan@post.kek.jp, on leave of absence from
          Theory Division, KEK, Tsukuba 305-0801, Japan},
    and Hai-Bo Yu\footnote{
  e-mail: hbyu@physics.umd.edu}}

\affiliation{Department of Physics,
 University of Maryland, College Park, MD 20742, USA}

\date{April, 2007}

\begin{abstract}

The minimal SUSY SO(10) GUT models with {\bf 10}, {\bf 126} and
{\bf 210} Higgs and only renormalizable couplings has been shown 
to provide a simple way to understand the neutrino mixings as well as the
ratio $\Delta m^2_\odot/\Delta m^2_A$ in terms of quark mixing
parameter $\theta_{Cabibbo}$, provided neutrino masses are
described by type II seesaw formula. However, in this minimal
picture, it is impossible to realize type II dominance with
renormalizable couplings in 4-dimensions. We show that this
problem can be cured by embedding this model into a warped
5-dimensional space time with warping between the Planck and the
GUT scale, where
 both type II as well as mixed seesaw formulae can be realized in a
natural manner without expanding the Higgs sector.
These models also avoid the possible problem of threshold effects
associated with large Higgs representations
since the theory above the GUT scale is now strongly coupled.
\end{abstract}
\maketitle
%\bigskip
%\bigskip

\section{Introduction}
Understanding neutrino masses and mixings has been a major challenge
to particle theorists. Many approaches have been proposed \cite{ms}.
While there is no consensus on the right final solution, some
important clues are emerging on which there appears a large degree
of agreement among theorists. If neutrino is a Majorana particle, then
seesaw mechanism\cite{seesaw} for understanding the origin
of its mass seems to have a strong appeal.
 The ingredients of this mechanism
are: (i) $m_\nu$ is related to B-L symmetry breaking, implying that
physics beyond the standard model must have this symmetry; B-L
 most likely is a local symmetry: (ii) secondly, it is also
possible that the breaking of this symmetry takes place at a
high scale by the Majorana mass of the right handed neutrinos which then
provides a natural way to understand the smallness of the neutrino
masses for natural values of parameters in the theory. A
theoretical support for this kind of scenario comes
from the observation that grand unified theories based on the SO(10)
group \cite{fm} automatically incorporate both the right handed
neutrinos into its spinor multiplets as well as the local B-L
symmetry as part of the gauge group and in most minimal ways of
symmetry breaking coupling constant unification requirement puts the
B-L symmetry breaking scale (and hence the right handed neutrino
mass) close  to the GUT scale of $10^{16}$ GeV, so that a high
seesaw scale close to GUT scale required for understanding
atmospheric neutrino observations becomes easier to understand.

The present paper addresses an important aspect of embedding
 the seesaw mechanism in a minimal SUSY SO(10) model. We focus on
SO(10) models with {\bf 126} Higgs field breaking B-L gauge symmetry
\cite{bm,nobu,vissani,goh, Dutta:2004wv} rather than the {\bf 16}
Higgs \cite{ab} since in the first case both R-parity symmetry of
MSSM and predictivity for neutrinos arise without imposing any extra
symmetries. We will discuss the class of models which we call
minimal SO(10) models because of the Higgs content of {\bf 10},
${\bf 126}\oplus \overline{\bf 126}$ and {\bf 210} and matter
content in three {\bf 16} spinors \cite{aulakh}.  In \cite{bm} and
several subsequent papers \cite{nobu}, the neutrino mass discussion
in this model was carried out using only the type I seesaw formula. But 
as is now
well known, there are two contributions to the seesaw formula
\cite{seesaw2} in left-right symmetric as well as SO(10) models i.e.
\begin{eqnarray}
{\cal M}_\nu~=~fv_L-M^T_D(fv)^{-1}M_D .
\end{eqnarray}
When the second term dominates, it is called type I seesaw whereas
when the first one dominates, it is called type II seesaw. The
advantage of the type II seesaw formula in understanding large
atmospheric neutrino mixings in a two generations minimal SO(10) model
was first observed in Ref.~\cite{vissani}. It was
subsequently shown \cite{goh} that the same scenario can help to
explain the large solar as well as small reactor mixing angle
$\theta_{13}$ bringing these models to
the mainstream of neutrino phenomenology. Other detailed questions
in the model such as CP violation \cite{CP}, proton decay \cite{pdecay} 
as well
as symmetry breaking \cite{sym} have since been discussed. Because
of predictivity in the neutrino sector while keeping the rest of
fermion mass phenomenology in agreement with observations as well
as general economy of the Higgs sector, these minimal models have
become very attractive and are in fact in a better footing than
SU(5) models were in the early 80's, with serious attention being
paid to them. One must therefore examine to what extent the model
parameters needed for the neutrino predictions can be naturally
obtained. It is this aspect of the models that we address in this
paper.

Since in the minimal SO(10) model, GUT symmetry relates the Dirac masses
of the neutrinos  to the up quark masses, one can ask for a more
quantitative understanding of the seesaw formula. For example, the
 atmospheric neutrino mass difference square $\Delta m^2_A\sim 0.0025$
eV$^2$ requires that at least one of the right handed neutrinos has
a mass around $10^{14}$ GeV, if one uses the type I seesaw formula
for neutrino masses. This is much less than the GUT scale which
determines the B-L breaking and therefore implies a fine tuning of
some Yukawa couplings. In the context of minimal SO(10) models, it
in fact turns out that fitting charged fermion masses also requires
a Yukawa coupling suppressed to that level\cite{goh}. Therefore they
go together and clearly, it will be important to understand this
mini-fine tuning from a more fundamental point of view\footnote{In
SO(10) models that use {\bf 16} Higgs fields \cite{ab}, since the
right handed neutrino Majorana masses come from higher dimensional
operators, this factor of 100 suppression comes from $M_{GUT}/M_{P}$
and hence easier to understand however further details of mixings
require new symmetries.}.

In this paper we concern ourselves with minimal SUSY SO(10) models
that use type II seesaw where a different
fine tuning becomes essential. The the magnitude of the type II
seesaw contribution to neutrino masses is given by
$f\frac{v^2_{wk}}{M_T}$ where $M_T$ is the B-L=2, SU(2)$_L$ triplet
mass and for $f\sim 1$, one needs $M_T\sim 10^{14}$ GeV whereas for
$f\sim 0.01$ as may be required by charged fermion fitting, we need
$M_T\sim 10^{12}$ GeV \footnote{Note that in non-SUSY SO(10) models,
there is an additional enhancement factor in the type II seesaw of
the form $M_{GUT}/M_T$ making the fine tuning problem less severe.
However such enhancement is absent in supersymmetric theories
\cite{rossi}.}. Since $M_T$ is related to $M_{GUT}$, the discrepancy
between them must be explained
% \footnote{Incidentally, in generic
%SO(10) models with {\bf 16}-Higgs breaking SO(10), type II
%contributions are naturally small compared to type I terms.}.
 An additional challenge for this class of models is that for type II
term to dominate, one must not only have the first term dominate in
Eq.~(1) but the second term must also be simultaneously smaller. In
the language of SU(5) submultiplets in the {\bf 126} field, $M_T$
must be the mass of the {\bf 15} sub-multiplet.

The problem in understanding type II dominance was discussed
 in Ref.~\cite{gmn} where it was shown that the requirements given
above for type II dominance cannot be satisfied
in the minimal four dimensional SUSY SO(10)
model with  ${\bf 10}\oplus
{\bf 126}\oplus {\bf 210}$ Higgs fields. The reason is that at high
scale there
are only four parameters in the superpotential and constraints of
supersymmetry imply that the triplet mass must be at the GUT scale,
making then type II term subdominant. This calls into question the
viability of the minimal models. The solution to this suggested in
\cite{gmn} was that the model be extended to include a {\bf 54}-dim.
Higgs field, in which case one can fine tune parameters to get a lower
triplet mass while at the same time suppressing the type I term.
 Since {\bf 54} Higgs does not couple
to matter fields, it does not affect the discussion of fermion masses and
mixings.

In this paper, we propose a different way to solve these fine tuning
problems without adding extra Higgs fields but rather by embedding
the minimal model into a warped 5-dimensional space time with
warping between the Planck scale and the GUT scale and with all
fields of the model in the bulk. We call this ``mini-warping'' since
the warp factor required here is $\omega\equiv M_{GUT}/M_P \sim
10^{-2}$ rather than the usual $m_W/M_P$ as in canonical
Randall-Sundrum (RS) models. Two things happen in such models if the
gauge group and other fields are in the bulk: (i) all mass
parameters in the IR brane are suppressed by $\omega$ and (ii)
depending on bulk mass and the gauge charge, there may be additional
suppression factors \cite{kita}. A combination of these two factors
provides a new way to resolve some of the fine tuning problems in
these models.

An initial application of this idea to understand type I seesaw in
minimal SO(10) has recently been discussed  by Fukuyama, Kikuchi and
Okada \cite{fuku} where it was shown how the smallness of the
right-handed neutrino mass can be understood as a consequence of
mini-warping. In the present paper, we show that mini-warping can also 
help to explain type II dominance of the seesaw formula. Unlike the case
of type I seesaw dominance, type II case involves a lot of
subtle issues such as the magnitude of the GUT scale, structure of
the MSSM doublets in terms of the GUT Higgs multiplets etc. and is
highly nontrivial due to interconnections between various terms in
the superpotential. We have however succeeded in finding an example
where this happens. This is the subject of this paper. The
significance of our result is that it restores the type II dominated
minimal SUSY SO(10) into a viable model.

The paper is organized as follows: in Sec. II we discuss the basic
ingredients of the approach; in Sec. III, we discuss the minimal
SO(10) and show how type II seesaw arises naturally without extra
Higgs fields; we discuss some implications of the model in Sec. IV.

%%%%%%%%%%%%%%%%%%%%%%%%%%%%%%%%%%%%%%%%%%%%%%%%%%%%
\section{Basic ingredients of a mini-warped model}
%%%%%%%%%%%%%%%%%%%%%%%%%%%%%%%%%%%%%%%%%%%%%%%%%%%%
 Our basic approach consists of embedding the minimal SO(10) model in the
warped five dimensional brane world scenario \cite{RS}
with warping between the Planck scale to the GUT scale.
The fifth dimension is compactified on the orbifold $S^1/Z_2$
 with two branes, ultraviolet (UV) and infrared (IR),
 located on the two orbifold fixed points.
As in the RS model, we use the warped metric \cite{RS},
\begin{eqnarray}
 d s^2 = e^{-2 k r_c |y|} \eta_{\mu \nu} d x^{\mu} d x^{\nu}
 - r_c^2 d y^2 \; ,
\end{eqnarray}
 with $-\pi\leq y\leq\pi$ and $\eta_{\mu\nu}= (+,-,-,-)$. In the above
expression, $k$ is the AdS curvature, and
 $r_c$ and $y$ are the radius and the angle of $S^1$, respectively.
As is well known, five dimensional $N=1$ SUSY corresponds to $N=2$
SUSY in four dimensions.
We can therefore write the 5-D superfields in terms of $N=2$
4-D multiplets. The process of compactification leads to $N=1$ SUSY on
the brane as well as in 4-D.

The Lagrangian for a generic $U(1)$ gauge theory with matter and
Higgs fields in the bulk can be written in terms of 4-D $N=1$
superfields as \cite{SUSYL}:
\begin{eqnarray}
{\cal L} &=& \int dy \left\{
\int d^4 \theta \; r_c \; e^{- 2 k r_c |y|}
 \left(
 H^{\dagger}_i  e^{- Q_i V} H_i  + H^{c}_i e^{Q_i V} H^{c \dagger}_i
 \right) \right. \nonumber \\
&+&
\left.
\int d^2 \theta e^{-3 k r_c |y|}
 H^{c}_i \left[
  \partial_{y} - \left( 1 + C_i \right) k r_c \epsilon(y)
 - Q_i \frac{\chi}{\sqrt{2}}  \right]
  H_i  +h.c. \right \} \; ,
\label{bulkL}
\end{eqnarray}
where $C_i$ is a dimensionless (bulk mass) parameter,
$\epsilon(y)=y/|y|$ is the step function,
 $H_i, ~H^c_i$ is the hypermultiplet with the charge $Q_i$
 under the gauge group, and
\bea
V &=& - \theta \sigma^\mu \bar{\theta} A_\mu
    -i \bar{\theta}^2 \theta \lambda_1  +i \theta^2 \bar{\theta}
\bar{\lambda}_1
    + \frac{1}{2} \theta^2 \bar{\theta}^2 D \; ,
\nonumber \\
\chi &=&  \frac{1}{\sqrt{2}}(\Sigma + i A_5) +\sqrt{2} \theta\lambda_2
     + \theta^2 F \; ,
\eea
 are the vector multiplet and the adjoint chiral multiplets,
 which form an $N=2$ SUSY gauge multiplet.
 $Z_2$ parity for $H_i$ and $V$ is assigned as even,
 while odd for $H^c_i$ and $\chi$.
This technique is easily generalized to the case of SO(10) model.
The point to emphasize is that in RS models,
 the mass scale of the IR brane is warped down
 by the warp factor \cite{RS}, $ \omega = e^{-k r_c \pi}$,
 in effective four dimensional theory.
If we take the cutoff of the original five dimensional
theory and the AdS curvature as $M_5 \simeq k \simeq M_P$,
the four dimensional (reduced) Planck mass, the cutoff scale in the IR brane
is $\Lambda_{IR}= \omega M_P$. In our case, we choose the warp factor
to be such that $ M_{GUT}= \Lambda_{IR}=\omega M_P$.
In the IR brane, the theory becomes non-perturbative above this scale
so that the question of large threshold corrections becomes moot.

Let us now assume that the gauge symmetry is broken down
and the adjoint chiral multiplet $\chi$ develops a VEV.
Since its $Z_2$ parity is odd, the VEV has to take the form,
\bea
\left<\Sigma \right> = 2 \alpha k r_c  \epsilon(y) .
\label{adjVEV}
\eea
In this case, the zero mode wave function of $H_i$
 satisfies the following equation of motion:
\bea
\left[\partial_y -
 \left(1+ C_i + Q_i \alpha \right) k r_c \epsilon(y) \right]H_i =0
\eea
which yields
\bea
H_i = \frac{1}{\sqrt{N_i}}
 e^{ (1 + C_i + Q_i \alpha) kr_c |y|} \; h_i(x^\mu) \; ,
\eea
where $h_i(x^\mu)$ is the chiral multiplet in four dimensions.
Here, $N_i$ is a normalization constant which ensures
that the kinetic term is canonically normalized. We have
\be
\frac{1}{N_i}
 =  \frac{2 (C_i + Q_i \alpha) k }
{e^{2 (C_i + Q_i\alpha) k r_c \pi}-1} \; .
\ee
There are now two typical cases to consider:

(i) if $e^{ (C_i + Q_i \alpha ) k r_c \pi}  \gg 1$,
the wave functions at $y=0$ and $y=\pi$ are, respectively, given by
\bea
H_i(y=0) &\simeq &
\sqrt{ 2(C_i + Q_i \alpha ) k } \; \omega^{C_i+Q_i \alpha} \; h(x^\mu).
\nonumber \\
H(y=\pi) &\simeq &
\sqrt{ 2(C_i + Q_i \alpha ) k } \; \omega^{-1} \; h(x^\mu).
\eea

(ii) whereas for $ e^{ (C_i + Q_i \alpha ) k r_c \pi}  \ll 1$,
the wave functions are
\bea
H(y=0) &\simeq &
\sqrt{- 2(C_i + Q_i \alpha ) k } \; h(x^\mu),
\nonumber \\
H_i(y=\pi) &\simeq &
\sqrt{- 2(C_i + Q_i \alpha ) k } \; \omega^{-(C_i+Q_i \alpha)}
 \omega^{-1} \; h(x^\mu)
\eea

\noindent
In case (i), the wave function is localized around the IR brane
 while around the UV brane in case (ii).
These non-trivial wave function profiles lead to important effects,
namely suppression of couplings and masses,
in effective four dimensional theory.

To see this, let us consider Yukawa couplings on the IR and UV branes
 for three bulk hypermultiplets:
\bea
{\cal L}_Y &=&
 \int d^2 \theta \omega^3
 \frac{Y_1}{M_5^{3/2}} H_i(y=\pi) H_j(y=\pi) H_k(y=\pi)
 \nonumber \\
&+&
 \int d^2 \theta
 \frac{Y_2}{M_5^{3/2}} H_i(y=0) H_j(y=0) H_k(y=0) + h.c. ,
\label{IR-UV-Yukawa}
\eea
where $Q_i + Q_j + Q_k=0 $ has been assumed for the U(1) gauge
invariance, and $Y_1$ and $Y_2$ are independent Yukawa coupling
constants on the IR and UV branes, respectively.
When all the bulk fields are localized around the IR brane
 ($C_{i,j,k}+Q_{i,j, k} \alpha > 0$),
we obtain the Yukawa coupling constant in effective four dimensional theory as
\bea
Y_{4D} \sim Y_1 + Y_2 \omega^{C_i+Q_i \alpha}
 \omega^{C_j+Q_j \alpha} \omega^{C_k+Q_k \alpha}
 \sim Y_1 .
\eea
There is no suppression for the Yukawa coupling constant
 on the IR brane while the Yukawa coupling constant on the UV brane
 is very much suppressed by the small wave function overlapping.
A more non-trivial example is to assume $H_i$ is localized around
the UV brane ($C_i+Q_i \alpha < 0$)
and the others are localized around the IR brane
 ($C_{j,k}+Q_{j, k} \alpha > 0$).
This case leads to the effective Yukawa coupling constant as
\bea
Y_{4D} \sim Y_1 \omega^{-(C_i+Q_i \alpha)} +
 Y_2 \omega^{C_j+Q_j \alpha} \omega^{C_k + Q_k \alpha}.
\eea
Both of the coupling constants are suppressed according to
the wave function overlapping between each field.
Other cases are completely analogous and the effective Yukawa coupling
constants are suppressed or not suppressed according to
the wave function profiles.

Next let us consider mass terms on the IR and UV branes
for two bulk hypermultiplet such as
\bea
{\cal L}_m &=&
 \int d^2 \theta \; \omega^3  \;
  \frac{m_1}{M_5} H_a(y=\pi) H_b(y=\pi)
\nonumber \\
&+&
 \int d^2 \theta
  \frac{m_2}{M_5} H_a(y=0)   H_a(y=0)  + h.c.
\label{IR-UV-Mass}
\eea
Here two mass terms on the IR and UV branes have been
generally introduced.
If two bulk fields are localized around the IR brane
($C_{a,b}+Q_{a,b} \alpha >0 $),
we obtain the mass term in effective four dimensional theory as
\bea
 m_{4D} \sim m_1 + m_2 \omega.
\eea
Although there is no suppression due to the wave function profiles
in this case, the mass term on the IR brane is warped down.
This is the characteristic feature of RS models mentioned above.
More general cases are, again, analogous and we find that
suppression factors (in addition to the warp factor) appear
in the effective mass according to the wave function overlap.

In the next section, we apply these results to explain the
naturalness of type I and type II seesaw in the minimal SO(10) model.
We will see that this goal can more or less be achieved except
we still need to do one fine tuning.

%%%%%%%%%%%%%%%%%%%%%%%%%%%%%%%%%%%%%%%%%%%%%%%%%%%%%%%%%%%%
\section{Relevant aspects of the minimal SUSY SO(10) model}
%%%%%%%%%%%%%%%%%%%%%%%%%%%%%%%%%%%%%%%%%%%%%%%%%%%%%%%%%%%%
In order to apply the discussion of the previous section to the
minimal SO(10) model, we provide a brief reminder of the salient
aspects of these models. All the couplings and mass parameters in
this model refer to four dimensions and we omit the superscript 4D
for all of them for simplicity. As long as we allow only
renormalizable couplings, the model has only two Yukawa coupling
matrices: (i) $h$ for the {\bf 10} Higgs and (ii) $f$ for the {\bf
126} Higgs. SO(10) has the property that the Yukawa couplings
involving the {\bf 10} and {\bf 126} Higgs representations are
symmetric. Therefore if we assume that CP violation arises from
other sectors of the theory (e.g. squark masses) and work in a
basis where one of these two sets of Yukawa coupling matrices is
diagonal, then there are only nine parameters describing the
Yukawa couplings. Noting the fact that the {\bf 45} and ${\bf
\bar{5}}$ SU(5)-submultiplets of ${\bf\overline{126}}$ has a pair
of standard model doublets in addition to the {\bf 5} and
$\bar{\bf 5}$ multiplets of ${\bf 10}$ that contributes to charged
fermion masses, one can write the quark and lepton mass matrices
as follows \cite{bm}:
\begin{eqnarray}
\nonumber M_u &=& h \kappa_u + f v_u \\\nonumber M_d &=& h \kappa_d + f v_d \\
\nonumber M_\ell &=& h \kappa_d -3 f v_d \\ M_{D}&=& h
\kappa_u -3 f v_u,
\end{eqnarray}
where $\kappa_{u,d}$ are the VEVs of the up and down standard model
type Higgs fields in the {\bf 10} multiplet and $v_{u,d}$ are the
corresponding VEVs for the same doublets in {\bf 126}. This gives 13
parameters describing the fermion masses and mixings (for both
leptons and quarks). If we input six quark masses, three lepton
masses and three quark mixing angles and weak scale, these are a
total of 13 parameters and all parameters are now determined.  Thus
all parameters of the model that go into fermion masses are
determined. The neutrino sector therefore has no free parameters
except an two overall scales ($v_L$ and $v_R$) as we see below:
\begin{eqnarray}
{\cal M}_\nu~=~2fv_L-M^T_D(2fv_R)^{-1}M_D
\end{eqnarray}
If type I or type II  seesaw dominates, except for an overall scale,
all the rest of the parameters of the neutrino mass matrix are
predicted. The problem addressed in this paper is to what extent one
can understand the naturalness of parameters that make either type I
or type II dominate.  As noted earlier, a simple understanding of
the large neutrino mixings \cite{vissani,goh} as well as an
explanation of the value of $\sqrt{\frac{\Delta m^2_\odot}{\Delta
m^2_A}}$ as being of order of the Cabibbo angle comes about in the
case of type II dominance.

 When one tries to understand CKM CP
violation in these models, it is useful to extend it by the
inclusion of a {\bf 120} Higgs field that couples to SM fermions
\cite{dutta}. We omit the {\bf 120} field from our considerations
since our main point is not affected by this.

To see what fine tunings are needed to make type II seesaw
dominate, let us write down the superpotential for the 4-D SUSY
SO(10) model that we are discussing. Denoting the {\bf 126} fields
by $\Sigma$, and {\bf 210} ones by $\Phi$, we have
\begin{eqnarray}
W~=~  M^{4D}_\Sigma \Sigma\bar{\Sigma}~+M^{4D}_\Phi \Phi^2 +
\lambda^{4D}_1 \Sigma\bar{\Sigma}\Phi +\lambda^{4D}_2 \Phi^3
\end{eqnarray}
where we have used the superscript 4-D to denote that this is a
4-D theory. It is helpful to write down the SU(5)$\times$U(1)$_X$
sub-multiplets
 of the various SO(10) multiplets used here:
\begin{eqnarray}
\nonumber{\bf 210}~&=&~{\bf 1}_0 \oplus {\bf 5}_{-8}\oplus {\bf
\overline{5}}_8 \oplus {\bf 10}_{4} \oplus {\bf
\overline{10}}_{-4}\oplus {\bf 24}_0
\oplus {\bf 75}_0 \oplus {\bf 40}_{-4}\oplus {\bf \overline{40}}_4,\\
\nonumber {\bf 126}~&=&~{\bf 1}_{-10}\oplus {\bf
\overline{5}}_{-2}\oplus {\bf 10}_{-6}\oplus {\bf
\overline{15}}_{+6}\oplus {\bf 45}_2\oplus {\bf
\overline{50}}_{-2},\\
{\bf 10}~&=&~{\bf 5_{2}}\oplus{\bf\overline{5}_{-2}}.
\end{eqnarray}
And the decomposition of matter field ${\bf 16}$ is
\begin{eqnarray}
{\bf 16}={\bf 1_{-5}}\oplus{\bf\overline{5}_{3}}\oplus{\bf
10_{-1}}.
\end{eqnarray}

\begin{figure}[h!]
\includegraphics[scale=0.9]{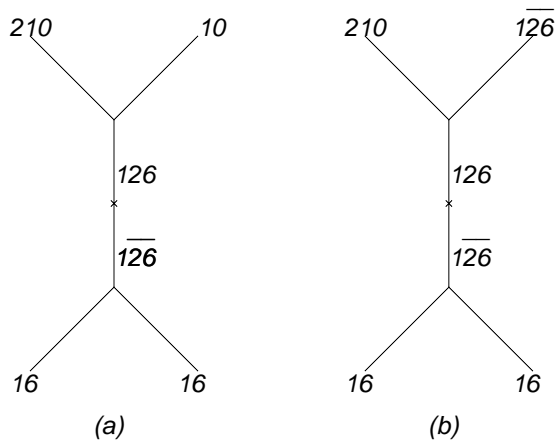}
\caption{Supergraph for type II seesaw.}
\end{figure}
The supergraph responsible for type II seesaw term is given in Fig.~1.
An inspection of this graph reveals that the following conditions
must be satisfied for the type II seesaw to be important for
neutrino mass discussion:

(i) $M_{15}\sim f\;  10^{-2} M_{GUT}$;

(ii) coupling ${\bf\overline{15}} \cdot {\bf 5}\cdot {\bf
5}\subset{\bf 210\cdot126\cdot10}$ or $\overline{\bf 15} \cdot {\bf
5} \cdot {\bf 5}\subset{\bf 210\cdot126\cdot\overline{126}}$ must
not be suppressed and be of order one.

We will show in the next section how we can have an understanding of
these two conditions within a mini-warped model using the technique
outlined in Sec.~II.

%%%%%%%%%%%%%%%%%%%%%%%%%%%%%%%%%%%%%%%%%%%%%%%%%%%%
\section{Minimal SO(10) theory in five dimensions}
%%%%%%%%%%%%%%%%%%%%%%%%%%%%%%%%%%%%%%%%%%%%%%%%%%%%
We take N=1 SUSY SO(10) model in five dimensions and put all the fields
(matter as well as Higgs) in the bulk with different bulk mass terms
for different fields. Note that all fields are paired with its complex
conjugate field so that the bulk mass terms are allowed
by gauge invariance and supersymmetry.
Note that these mass terms play the role of a parameter
describing the wave function profile of the field and
are not the mass terms of 4-D theory.

We put the interaction terms on both IR and UV branes. Both $\bf
126$ and $ \bf 10$ mass terms on the IR brane, and the mass term of
$\bf 210$ on the UV brane. The relevant part of the Lagrangian can
be written as ${\cal L}=\int d^2 \theta W_{IR}
       +  \int d^2 \theta W_{UV} + h.c. $, where
\begin{eqnarray}
W_{IR} &=& %\int d^2\theta
\omega^3
\left[
 %\frac{M_{\Phi}}{M_5}\Phi^2+
\frac{M_{\Sigma}}{M_5}\Sigma\overline{\Sigma}+
\frac{M_H}{M_5}H^2+\frac{\lambda_1}{M^{3/2}_5} \Phi^3
%\right.
% \nonumber \\
%
%&+&  \left.
+
\frac{\eta_1}{M^{3/2}_5}\Phi\Sigma\overline\Sigma+
\frac{1}{M^{3/2}_5}\Phi H (\alpha_1\Sigma+{\overline\alpha_1}
{\overline\Sigma})
\right]_{y=\pi}, %+h.c.
\nonumber \\
W_{UV} &=&
 %\int d^2\theta
\left[
\frac{M_{\Phi}}{M_5}\Phi^2+\frac{\lambda_2}{M^{3/2}_5}
 \Phi^3+\frac{\eta_2}{M^{3/2}_5}\Phi\Sigma\overline\Sigma
 +\frac{1}{M^{3/2}_5}\Phi H
 (\alpha_2\Sigma+{\overline\alpha_2}{\overline\Sigma})
\right]_{y=0} .
%+h.c.
\label{HiggsW}
\end{eqnarray}
%
%Naturalness requires that the couplings on the UV and IR
%branes are of same order.
%
Suppose that the couplings on the UV and IR branes are of the same order.

Now we assume that the adjoint chiral multiplet of U(1)$_X$
 has non-zero VEV as in Eq.~(\ref{adjVEV}) \footnote{
Since $Z_2$ parity for this field is assigned as odd,
 the non-zero VEV leads to the Fayet-Iliopoulos D-terms
 localized on both the UV and IR branes \cite{D-term},
 which should be canceled to preserve SUSY.
For this purpose, we need to introduce new fields
on both branes by which the D-terms are compensated.
If such fields are in the same representations
 as matters or Higgs fields like ${\bf 16}$ or $\overline{\bf 126}$,
 we would need to impose some global symmetry to distinguish them.
}
and gives additional contributions to the bulk mass parameters
for the bulk fields.
In the following, we denote each chiral field of
SU(5)-submultiplets in $H_i$
as $H_{im}=(\Phi_{m},H_{m},\Sigma_{m},\overline\Sigma_{m})$,
where $m$ specifies the dimension of the submultiplets.
The zero mode solution of $H_{im}$ is described as
\begin{eqnarray}
H_{im}(x,y)=\kappa_{im}\sqrt{k}e^{kr_c|y|}
 e^{(C_i+\alpha Q_{im})kr_c|y|}h_{im}(x) ,
\end{eqnarray}
where
 $\kappa_{im}\equiv\sqrt{\frac{2(C_i+\alpha Q_{im})}{e^{2(C_i+\alpha
  Q_{im})kr_c\pi}-1}}$.
On the IR brane
 $H_{im}(x,\pi)=\kappa_{im}\sqrt{k}\omega^{-1}
 \omega^{-(C_i +\alpha Q_{im})}h_{im}(x)$ while
 $H_{im}(x,0)=\kappa_{im}\sqrt{k}h_{im}(x)$ on the UV brane.

We take $M_{\Sigma}$ and $M_{H}$ to be $\sim M_{P}$
 and $M_{\Phi}$ to be $\sim M_{GUT}$.
Because of the warp factor $\omega$,
 the 4-D effective masses of the IR brane
 are warped down to $\omega M_{P}\simeq M_{GUT}$.
Next note that
\begin{eqnarray}
e^{(C_i+\alpha Q_{im})k r_c \pi}&\gg& 1,  \;
 \kappa_{im}\simeq\sqrt{2(C_i+\alpha Q_{im})} \;
 \omega^{C_i+\alpha Q_{im}} \\
e^{(C_i+\alpha Q_{im}) kr_c \pi}&\ll& 1,  \;
 \kappa_{im}\simeq\sqrt{-2(C_i+\alpha Q_{im})}   \\
e^{(C_i+\alpha Q_{im}) kr_c \pi}&=& 1,  \;
 \kappa_{im}\simeq\sqrt{-\frac{1}{\ln\omega}}.
\end{eqnarray}
The extent of suppression of couplings and masses
in effective four dimensional theory are determined
by parameters $C_i$ and $\alpha$.
In this paper, we choose the parameters as listed in Tables.

%%%%%%%%%%%%%%%%%%%%%%%%%%%%%%%%%%%%%%%%%%%%%%%%%%%%%
\subsection{Masses of submultiplets of ${\bf 126}$}
%%%%%%%%%%%%%%%%%%%%%%%%%%%%%%%%%%%%%%%%%%%%%%%%%%%%%
As noted in Sec. III, one main problem for the minimal 4-D SO(10)
is that the SU(5)-submultiplets ${\bf 15,50}$ and ${\bf 45}$ have
the same mass $M_{\Sigma}$ (up to the Clebsch-Gordan (CG)
coefficients) \cite{gmn}. When we lower the ${\bf 15}$ Higgs mass
so as to obtain type II dominance, other Higgs fields accordingly
becomes light. As a result, gauge couplings blow up before they
unite at the GUT scale. As we show now, the situation is very
different in the mini-warped model.

Under the SU(5) decomposition, the mass term of the ${\bf 126}$ pair
 on the IR brane can be written as
\begin{eqnarray}
 \int d^2 \theta \omega^3 \left[
  \frac{M_{\Sigma}}{M_5} \Sigma\overline{\Sigma}  \right]_{y=\pi}
&\sim & \int d^2 \theta \;
 m_{\Sigma} \left[
 \epsilon_{\sigma0}\epsilon_{\overline{\sigma}0}\sigma_0\overline\sigma_0
+\epsilon_{\sigma15}\epsilon_{\overline{\sigma}15}\sigma_{15}\overline\sigma_{15}
+\epsilon_{\sigma10}\epsilon_{\overline{\sigma}10}\sigma_{10}\overline
\sigma_{10}
\right.  \nonumber \\
&+&
\left.
\epsilon_{\sigma50}\epsilon_{\overline{\sigma}50}\sigma_{50}
\overline\sigma_{50}+\epsilon_{\sigma45}
\epsilon_{\overline{\sigma}45}\sigma_{45}\overline\sigma_{45}
+\epsilon_{\sigma5}\epsilon_{\overline{\sigma}5}\sigma_5\overline\sigma_5
\right],
\label{mass15}
\end{eqnarray}
where $ m_{\Sigma}= \omega M_{\Sigma} \sim M_{GUT}$,
and $\epsilon_{im}\equiv\kappa_{im}\omega^{-(C_i+\alpha Q_{im})}$.
From Table \ref{higgs126} and Table \ref{higgs126bar},
 we have $\epsilon_{\sigma15}\sim\omega^{3/2}$ and
        $\epsilon_{\overline{\sigma}15}\sim 1 $,
therefore the mass of ${\bf 15}$ is suppressed by the factor
 $\omega^{3/2}$ and
 $M_{\bf 15} \sim \omega^{3/2} M_{GUT} \sim 10^{13}$ GeV.
On the other hand, we read
 $\epsilon_{\sigma50} = \epsilon_{\sigma50} \sim 1$,
 so the mass of ${\bf 50}$ is $\sim M_{GUT}$.
For ${\bf 45}$,
 $\epsilon_{\sigma45} \sim \omega^{1/2}$ and
 $\epsilon_{\overline{\sigma}45} \sim 1$,
 and its mass is $\sim \omega^{1/2}M_{GUT} \sim 10^{15}$ GeV.
In our mini-warped SO(10) model,
there is no mass degeneracy between these submultiplets.

This mass splitting also leaves gauge coupling unification of MSSM 
unchanged, since the submultiplets are all full SU(5) multiplets.
It is easy to check that the unified gauge coupling value at the GUT 
scale i.e. $\alpha_{GUT}\sim 0.2$ which
is in the perturbative regime  even though the ${\bf 15} \oplus 
\overline{\bf 15}$
multiplets with mass around $10^{13}$ GeV
and the ${\bf 45} \oplus \overline{\bf 45}$ multiplets
with mass around $10^{15}$ GeV are involved
into the gauge coupling running.

%%%%%%%%%%%%%%%%%%%%%%%%%%%%%%%%%
\subsection{Symmetry breaking}
%%%%%%%%%%%%%%%%%%%%%%%%%%%%%%%%%
Here we examine the realization of the SO(10) symmetry breaking.
Let us first see the SO(10) gauge symmetry breaking down to SU(5).
There are three SU(5) singlets:
one in $\bf 210$ and one in each of the $\bf 126$ pair
with non-zero B-L charge.
Since supersymmetry must remain unbroken all the way
down to the weak scale, F-flatness conditions determine
vacuum expectation values.
The relevant part in the superpotential in Eq.~(\ref{HiggsW})
is given by
\begin{eqnarray}
& &\int d^2\theta
\omega^3  \left[
 \frac{M_{\Sigma}}{M_5}\Sigma\overline{\Sigma}
 +\frac{\lambda_1}{M^{3/2}_5} \Phi^3
 + \frac{\eta_1}{M^{3/2}_5}\Phi\Sigma\overline\Sigma
\right]_{y=\pi}
+ \left[
\frac{M_{\Phi}}{M_5}\Phi^2+\frac{\lambda_2}{M^{3/2}_5}
 \Phi^3+\frac{\eta_2}{M^{3/2}_5}\Phi\Sigma\overline\Sigma
\right]_{y=0} \nonumber \\
&\supset &
m_{\Sigma} \epsilon_{\sigma0}\epsilon_{\overline{\sigma}0}
\sigma_0\overline\sigma_0
+ M_{\Phi} \kappa^2_{\phi0}\phi_0^2
+  (\lambda_1\epsilon^3_{\phi0} +   \lambda_2\kappa^3_{\phi0})\phi_0^3
+ (\eta_1{\epsilon_{\phi0}\epsilon_{\sigma0}\epsilon_{\overline\sigma0}}+
\eta_2\kappa_{\phi0}\kappa_{\sigma0}\kappa_{\overline\sigma0})
 \sigma_0 \overline\sigma_0 \phi_0
\nonumber \\
& \sim &
m_{\Sigma} \omega^{3/2} \sigma_0\overline\sigma_0
+ M_{\Phi} \phi_0^2
+ (\lambda_1 \omega^6 +   \lambda_2 ) \phi_0^3
+ (\eta_1 \omega^{5/2} + \eta_2 \omega^{7/2} )
 \sigma_0 \overline\sigma_0 \phi_0  .
\end{eqnarray}

F-flatness conditions for $\overline\sigma_0$ and $\phi_0$
lead to
\begin{eqnarray}
&&
 \sigma_0
\left[
  m_{\Sigma} \omega^{3/2} +
 (\eta_1 \omega^{5/2} + \eta_2 \omega^{7/2} ) \phi_0
\right] = 0,
\nonumber \\
&&
2 M_{\Phi} \phi_0 + 3 (\lambda_1 \omega^6 + \lambda_2 ) \phi_0^2
+ (\eta_1 \omega^{5/2} + \eta_2 \omega^{7/2} )
 \sigma_0 \overline\sigma_0  =0,
\end{eqnarray}
and the solutions are
\begin{eqnarray}
\left<\phi_0\right>
%= -\frac{m_{\Sigma}\omega^{3/2}}{(\eta_2\omega^{5/2}+\eta_1\omega^{7/2})}
\simeq-\frac{m_\Sigma}{\eta_2\omega},~
\left<\sigma_0\overline\sigma_0\right>  \simeq
-\frac{2 M_{\Phi} \left<\phi_0\right>}{\eta_2\omega^{5/2}}
\left(
1 + \frac{3\lambda_2 \left<\phi_0\right>}{2 M_{\Phi}}
\right).
\end{eqnarray}
SO(10) gauge symmetry is broken down to SU(5)$\times$U(1)$_X$ by
$\left<\phi_0\right>$ at the scale $m_\Sigma/(\eta_2 \omega)$. More
correctly, when we carefully consider the CG coefficients and
normalization of submultiplets of SO(10) under SU(5), we have an
extra factor $10$ accompanying with the coupling $\eta_2$
\cite{gmn}. Thus, if we take, for example, $\eta_2 \sim 4\pi$ this
symmetry breaking occurs around the GUT scale, $ \left<\phi_0\right>
\sim m_\Sigma/(10 \eta_2 \omega) \sim M_{GUT}$. On the other hand,
in order to arrange the B-L breaking scale to be around the GUT
scale, one needs to fine tune the coupling $\lambda_2$ to be
$\lambda_2\sim 1-\omega^{3/2}$.

Next we consider the SU(5) symmetry breaking by $\bf 24$ VEV.
The relevant superpotential is given by
\begin{eqnarray}
& &\int d^2\theta
\omega^3  \left[
  \frac{\lambda_1}{M^{3/2}_5} \Phi^3
\right]_{y=\pi}
+ \left[
\frac{M_{\Phi}}{M_5}\Phi^2+\frac{\lambda_2}{M^{3/2}_5} \Phi^3
\right]_{y=0} \nonumber \\
&\supset &
 M_{\Phi} \kappa^2_{\phi24} \phi_{24}^2
+ (\lambda_1 \epsilon_{\phi0} \epsilon^2_{\phi24}
+  \lambda_2 \kappa_{\phi0} \kappa^2_{\phi24})
\phi_0 \phi_{24}^2
+ (\lambda_1\epsilon^3_{\phi24} +   \lambda_2\kappa^3_{\phi24})
  \phi_{24}^3
\nonumber \\
& \sim &
M_{\Phi} \phi_{24}^2
+ (\lambda_1 \omega^6 +   \lambda_2 ) \phi_0 \phi_{24}^2
+ (\lambda_1 \omega^6 +   \lambda_2 ) \phi_{24}^3 .
\end{eqnarray}
Through the F-flatness condition for $\phi_{24}$, we obtain
\begin{eqnarray}
\left<\phi_{24}\right> \sim
- \frac{M_{\phi} + \lambda_2\left<\phi_0\right>}{\lambda_2}
\sim M_{GUT}.
\end{eqnarray}

Once $\phi_0$ gets the VEV, a new contribution appears
to the mass of $\bf 15$ through the superpotential,
\bea
&&
\int d^2 \theta \; \omega^3
 \left[
  \frac{\eta_{1}}{M^{3/2}_5} \Phi \Sigma \overline\Sigma
 \right]_{y=\pi}
+
 \left[
  \frac{\eta_{2}}{M^{3/2}_5} \Phi \Sigma \overline\Sigma
 \right]_{y=0}
\nonumber \\
&\supset &
\left[
\eta_1 \epsilon_{\phi0} \epsilon_{\sigma15}
\epsilon_{\overline{\sigma}15}+
\eta_2 \kappa_{\phi0} \kappa_{\sigma15} \kappa_{\overline{\sigma}15}
\right]
 \left< \phi_0 \right> \sigma_{15}\overline\sigma_{15}
\nonumber \\
&\sim &
\left[
\eta_1 \omega^{7/2} + \eta_2 \omega^{5/2} \right]
 \left< \phi_0 \right> \sigma_{15}\overline\sigma_{15} .
 \label{extra15mass1}
\eea
%
%Using the assignments given in TABLE, the suppression facts are
%$\epsilon_{\phi0}\sim\omega^2$,$\kappa_{\phi0}\sim{\cal O(\rm
%1)},\kappa_{\sigma15}\sim{\cal O(\rm 1)}$ and $
%\kappa_{\overline{\sigma}15}\sim{\omega^{5/2}}$, therefore the
%overall mass contribution from Eq.~(\ref{extra15mass1}) is
%$\sim\omega^{5/2}\eta_2\left<\phi_0\right> \simeq
%\omega^{3/2}M_{GUT}$.
%
Substituting the above $\left< \phi_0 \right>$ into this formula,
we find the additional contribution of order $ \omega^{3/2}M_{GUT}$,
that is the same order as the one from the tree level mass term
in Eq.~(\ref{mass15}).

%%%%%%%%%%%%%%%%%%%%%%%%%%%%%%%%%%%%%%%%%%%%%%%%
\section{Neutrino mass and Type II dominance}
%%%%%%%%%%%%%%%%%%%%%%%%%%%%%%%%%%%%%%%%%%%%%%%%
In this section we show how type II dominance emerges in our model.
Yukawa couplings on both the IR and UV branes are given by
\begin{eqnarray}
\int d^2\theta\omega^3
\left[
 \frac{f_{1ab}}{M^{3/2}_5}\Psi_a\Psi_b\overline{\Sigma}+
 \frac{h_{1ab}}{M^{3/2}_5}\Psi_a\Psi_bH
\right]_{y=\pi}
+ \left[
 \frac{f_{2ab}}{M^{3/2}_5}\Psi_a\Psi_b\overline{\Sigma}+
 \frac{h_{2ab}}{M^{3/2}_5}\Psi_a\Psi_bH
\right]_{y=0},
\end{eqnarray}
where $\Psi_a$ is the ${\bf 16}$ matter field
of the $a$-th generation ($a=1,2,3$).

We first consider the Yukawa coupling for
 $\bf\overline 5\cdot\overline5\cdot 15$,
 which is extracted as
\begin{eqnarray}
\left[
 f_{1ab}\epsilon^2_{\psi\overline5}\epsilon_{\overline\sigma15}
+
 f_{2ab}\kappa^2_{\psi\overline5}\kappa_{\overline\sigma15}
\right]
 \psi_{\overline5}\psi_{\overline5}\overline\sigma_{15}
\sim
\left[ f_{1ab} \omega^{1/2} +
       f_{2ab} \omega^{5/2} \right]
 \psi_{\overline5}\psi_{\overline5}\overline\sigma_{15} .
\end{eqnarray}
Now the effective Yukawa coupling in 4-D is found to be
 $\sim  f_{1ab} \omega^{1/2} $.

In Fig.~1, there are two vertexes between Higgs fields
 involved in type II seesaw formulas,
 ${\bf 210\cdot126\cdot10}$ or
 ${\bf 210\cdot126\cdot\overline{126}}$.
%
%which includes
%$\phi_5h_5\sigma_{\overline{15}}$ and
%$\phi_5\overline{\sigma}_5\sigma_{\overline{15}}$ respectively.
%
 From the superpotential in Eq.~(\ref{HiggsW})
 the vertex in Fig.~1(a) can be read off as
\begin{eqnarray}
\left[
\alpha_{1}\epsilon_{\phi5}\epsilon_{h5}
 \epsilon_{\sigma\overline{15}}+
\alpha_2\kappa_{\phi5}\kappa_{h5}\kappa_{\sigma\overline{15}}
 \right]
\phi_5h_5\sigma_{\overline{15}}.
\end{eqnarray}
From Tables,
 $\epsilon_{\phi5} \sim \epsilon_{h5} \sim 1$,
 $\epsilon_{\sigma\overline{15}} \sim \omega^{3/2}$,
 and $\kappa_{\phi5} \sim \kappa_{h5} \sim
 \kappa_{\sigma\overline{15}} \sim 1$,
 so that we have the coupling
 $\sim \alpha_2 \phi_5 h_5 \sigma_{\overline{15}}$
 un-suppressed.
On the other hand, for the vertex in Fig.~1(b),
 we have
\begin{eqnarray}
\left[
\eta_{1}\epsilon_{\phi5}\epsilon_{\overline{\sigma}5}
\epsilon_{\sigma\overline{15}}+\eta_{2}\kappa_{\phi5}
\kappa_{\overline{\sigma}5}\kappa_{\sigma\overline{15}}
\right]
 \phi_5\overline{\sigma}_5\sigma_{\overline{15}} .
\end{eqnarray}
This contribution is negligible compared to the previous one,
 since $\epsilon_{\overline{\sigma}5}\sim 1$
 and $\kappa_{\overline{\sigma}5} \sim \omega^{1/2}$.

We are now ready to estimate the relative magnitudes of the two 
different seesaw contributions to neutrino mass in our model.  For this 
purpose, we note that in terms of the original SO(10) Yukawa couplings 
 the $f_1 {\bf 16}\cdot{\bf 16}\cdot{\bf \overline{126}}$, 
we can rewrite the seesaw formula as
\begin{eqnarray} 
{\cal M}_\nu~=~2f_1v_L-M^T_D(2f_1v_R)^{-1}M_D 
\end{eqnarray}
The magnitude of the neutrino mass 
from
 the Type II seesaw contribution is estimated as
\begin{eqnarray}
 M^{II}_{\nu} \simeq
 \frac{2(f_1)_{33} \omega^{1/2} v_{10} v_{210}
   \alpha_2}{M_{GUT}\omega^{3/2}},
\end{eqnarray}
where $v_{10,210}$ is the VEV of up-type Higgs doublets
 in ${\bf 10}$ and ${\bf 210}$.
If we take $(f_1)_{33} \sim 1$, $\alpha_2 \sim 0.5$
 and assume $v_{10} \simeq v_{210} \sim 100$ GeV,
 we arrive at the reasonable value
 for the atmospheric neutrino oscillation data,
 $M^{II}_\nu \simeq 0.05$ eV. Note however that $b-\tau$ unification as 
well as charge fermion fitting implies that ${(f_1)}_{33}\sim 
0.037$~\cite{Dutta:2004wv}.
In this case also one can get type II term to be $0.046$ eV if 
$\alpha_2\sim 4\pi$ and perturbative.

Next let us examine type I seesaw contribution.
The right-handed neutrino mass can be read as
\begin{eqnarray}
\left[
  f_{1ab}\epsilon^2_{\psi1}\epsilon_{\overline\sigma0}
 +f_{2ab}\kappa^2_{\psi1}\kappa_{\overline\sigma0}
\right] \left<\overline\sigma_{1}\right>
 \sim\omega^{3/2} f_{1ab} M_{GUT} .
\end{eqnarray}
Thus, the type I seesaw contribution is found to be
\begin{eqnarray}
M^{I}_{\nu} = M^T_D M_R^{-1} {M_D}
 \simeq \frac{m^2_{t} \omega^{1/2}}{2(f_1)_{33}M_{GUT}\omega^{3/2}},
\end{eqnarray}
where $m_t$ is top quark mass, and
we have used the natural relation $M_D \sim m_t$ in GUT models.
Using $m_t \sim 100$ GeV at the GUT scale,
the type I seesaw gives the contribution to
the neutrino mass as $M^{I}_{\nu} \simeq 0.025$ eV for ${(f_1)}_{33}\sim 
1$, which is already smaller than the type II seesaw contribution.
Again for the case of ${(f_1)}_{33}\sim 0.037$ obtained from charged 
fermion 
fitting in  Ref.~\cite{Dutta:2004wv},
even though the naive order of magnitude estimate  for $m_\nu$ from type 
I seesaw may appear to be large, full 
matrix effects from $M_D$ and $M_R$ indeed gives the desired neutrino 
masses.
For example, if we use  the explicit forms for the coupling matrices 
given in Ref.~\cite{Dutta:2004wv}, with $(f_1)_{33}\simeq 0.035$ 
using Eq. (V.7), we get the right order for $m_3$ even though 
 naive estimates would have suggested
$m_{\nu} \simeq 0.5$ eV.

% As $f_1$ becomes larger, type I 
%contribution reduces while type II contribution becomes larger.
%Note that in type II seesaw formula,
% the coupling $\alpha_2$ is a free parameter
% so that we can keep the type II contribution
% to be the same by changing $\alpha_2$.
%Therefore, according to a suitable parameter choice,
%we can realize all the cases:
%type I dominance, type II dominance, and mixed case.

%%%%%%%%%%%%%%%%%%%%%%
\section{Conclusion}
%%%%%%%%%%%%%%%%%%%%%%
In conclusion, we have shown that unlike the 4-dimensional minimal SUSY
SO(10) models where it is not possible to achieve type II
dominance of the seesaw formula, embedding into a mini-warped 5-D
space-time cures this problem  and leads to an effective 4-D
theory where either type II or mixed seesaw can dominate the neutrino 
mass. Thus the simple understanding of the large neutrino mixings
as well as the right solar mass difference square obtained in minimal 
SUSY SO(10) models is based on sound theoretical footing and no new 
Higgs fields need be added. We 
have also analyzed the symmetry breaking of SO(10) down to the standard
model in this framework and we found that to maintain the SU(5) and 
SO(10) scales at
$10^{16} $ GeV in this model, we need to fine tune only one parameters by 
a factor of $10^{-3}$. Note that in the minimal 4-D SO(10) model, we
could not even do any fine tuning to get the desired feature of type II 
dominance.  We have also checked that the SU(5) multiplets below
the GUT scale not only do not affect unification as expected but
they also keep the GUT couplings $\alpha_{GUT}\sim 0.2$ meaning
that one can use perturbation theory up to the GUT scale without
any problem.

%%%%%%%%%%%%%%%%%%%%%%%%%%
\acknowledgments
%%%%%%%%%%%%%%%%%%%%%%%%%%
The works of R.N.M. and H.B.Y. are supported by the National Science
Foundation Grant No. PHY-0354401.
The work of N.O. is supported in part by the Grant-in-Aid
 for Scientific Research from the Ministry of Education,
 Science and Culture of Japan (\#18740170).

%%%%%%%%%%%%%%%%%%%%%%%%%%%

\newpage

\begin{table}\label{fermion16}
\begin{tabular}{|c| c|} \hline
{\bf{16}} components & $C_{\rm \bf{16}}+\alpha Q_i$\\
\hline ${\rm\bf 1_{-5}}$&$7/4$\\
\hline ${\rm\bf 10_{-1}}$&$3/4$\\
\hline ${\rm\bf \overline{5}_{3}}$&$-1/4$\\
\hline
\end{tabular}
\caption{$C_{\rm\bf {16}}=1/2$ and $\alpha=-1/4$}
\end{table}

\begin{table}
\begin{tabular}{|c| c|} \hline
{\bf{10}} components & $C_{\rm \bf{10}}+\alpha Q_i$\\
\hline ${\rm\bf 5_{2}}$&$0$\\
\hline ${\rm\bf \overline{5}_{-2}}$&$1$\\
\hline
\end{tabular}
\caption{$C_{\rm\bf {10}}=1/2$ and $\alpha=-1/4$}
\label{higgs10}
\end{table}

\begin{table}
\begin{tabular}{|c| c|} \hline
{\bf{126}} components & $C_{\rm \bf{126}}+\alpha Q_i$\\
\hline ${\rm\bf 1_{-10}}$&$5/2$\\
\hline ${\rm\bf \overline5_{-2}}$&$1/2$\\
\hline ${\rm\bf {10}_{-6}}$&$3/2$\\
\hline ${\rm\bf {\overline {15}}_{6}}$&$-3/2$\\
\hline ${\rm\bf 45_{2}}$&$-1/2$\\
\hline ${\rm\bf \overline{50}_{-2}}$&$1/2$\\
\hline
\end{tabular}
\caption{$C_{\rm\bf {126}}=0$ and $\alpha=-1/4$}
\label{higgs126}
\end{table}

\begin{table}
\begin{tabular}{|c| c|} \hline
${\bf \overline{126}}$ components & $C_{\rm \bf{\overline{126}}}+\alpha Q_i$\\
\hline ${\rm\bf 1_{10}}$&$-3/2$\\
\hline ${\rm\bf 5_{2}}$&$1/2$\\
\hline ${\rm\bf {\overline{10}}_{6}}$&$-1/2$\\
\hline ${\rm\bf {{15}}_{-6}}$&$5/2$\\
\hline ${\rm\bf \overline{45}_{-2}}$&$3/2$\\
\hline ${\rm\bf {50}_{2}}$&$1/2$\\
\hline
\end{tabular}
\caption{$C_{\rm\bf {\overline{126}}}=1$ and $\alpha=-1/4$}
\label{higgs126bar}
\end{table}

\begin{table}
\begin{tabular}{|c| c|} \hline
${\bf 210}$ components & $C_{\rm \bf{210}}+\alpha Q_i$\\
\hline ${\rm\bf 1_{0}}$&$-2$\\
\hline ${\rm\bf 5_{-8}}$&$0$\\
\hline ${\rm\bf \overline{5}_{8}}$&$-4$\\
\hline ${\rm\bf 10_{4}}$&$-3$\\
\hline ${\rm\bf \overline{10}_{-4}}$&$-1$\\
\hline ${\rm\bf 24_{0}}$&$-2$\\
\hline ${\rm\bf \overline{40}_{4}}$&$-3$\\
\hline ${\rm\bf {40}_{-4}}$&$-1$\\
\hline ${\rm\bf 75_{0}}$&$-2$\\
\hline
\end{tabular}
\caption{$C_{\rm\bf {210}}=-2$ and $\alpha=-1/4$}
\label{higgs210}
\end{table}


\begin{thebibliography}{90}
\bibitem{ms} For a recent review and references, see R.~N.~Mohapatra and
A.~Y.~Smirnov, Ann.\ Rev.\ Nucl.\ Part.\ Sci.\  {\bf 56}, 569 (2006).


\bibitem{seesaw}  P. Minkowski, Phys. Lett. {\bf B 67}, 421 (1977);
 M. Gell-Mann, P. Ramond and R. Slansky, in {\it
Supergravity}, eds. D. Freedman {\it et al.} (North-Holland,
Amsterdam, 1980); T. Yanagida, in proc. KEK workshop, 1979
(unpublished); R. N. Mohapatra and G. Senjanovi\'c, Phys. Rev.
Lett. {\bf 44}, 912 (1980); S. L. Glashow, in {\it Proceedings of
1979 Cargese Summer Institute on Quarks and Leptons}, eds. M. Levy
{\it et al} , Plenum
 Press, New York, 1980, pp. 687-713.

\bibitem{fm}  H.~Fritzsch and P.~Minkowski,
  %``Unified Interactions Of Leptons And Hadrons,''
  Annals Phys.\  {\bf 93}, 193 (1975); H. Georgi, (1975).

\bibitem{bm} K.~S.~Babu and R.~N.~Mohapatra,
  %``Predictive Neutrino Spectrum In Minimal SO(10) Grand Unification,''
  Phys.\ Rev.\ Lett.\  {\bf 70}, 2845 (1993)

\bibitem{nobu}  T.~Fukuyama and N.~Okada,  JHEP {\bf 0211}, 011 (2002).

\bibitem{vissani} B.~Bajc, G.~Senjanovic and F.~Vissani,
  Phys.\ Rev.\ Lett.\  {\bf 90}, 051802 (2003).

\bibitem{goh}  H.~S.~Goh, R.~N.~Mohapatra and S.~P.~Ng,
  Phys.\ Lett.\ B {\bf 570}, 215 (2003); Phys.\ Rev.\ D {\bf 68}, 115008
(2003).

\bibitem{Dutta:2004wv}
  B.~Dutta, Y.~Mimura and R.~N.~Mohapatra,
  %``CKM CP violation in a minimal SO(10) model for neutrinos and its
  %implications,''
  Phys.\ Rev.\  D {\bf 69}, 115014 (2004)
  [arXiv:hep-ph/0402113].
  %%CITATION = PHRVA,D69,115014;%%

\bibitem{ab} C.~H.~Albright, K.~S.~Babu and S.~M.~Barr,
Phys.\ Rev.\ Lett.\  {\bf 81}, 1167 (1998)
[arXiv:hep-ph/9802314];
K.~S.~Babu, J.~C.~Pati and F.~Wilczek,
Nucl.\ Phys.\ B {\bf 566}, 33 (2000)
[arXiv:hep-ph/9812538];
C.~H.~Albright and S.~M.~Barr,
Phys.\ Rev.\ Lett.\  {\bf 85}, 244 (2000)
[arXiv:hep-ph/0002155];
T.~Blazek, R.~Dermisek and S.~Raby,
Phys.\ Rev.\ D {\bf 65}, 115004 (2002)
[arXiv:hep-ph/0201081]; X.~d.~Ji, Y.~c.~Li and R.~N.~Mohapatra,
  Phys.\ Lett.\  B {\bf 633}, 755 (2006).


\bibitem{aulakh} C.~S.~Aulakh and R.~N.~Mohapatra,
  Phys.\ Rev.\  D {\bf 28}, 217 (1983); T.~E.~Clark, T.~K.~Kuo and
N.~Nakagawa, Phys.\ Lett.\  B {\bf 115}, 26 (1982).


\bibitem{seesaw2}  G. Lazarides, Q. Shafi and C. Wetterich,
Nucl.Phys.{\bf B181}, 287 (1981); R. N. Mohapatra and G.
Senjanovi\'c, Phys. Rev. {\bf D 23}, 165 (1981).

\bibitem{rossi} A.~Rossi,  Phys.\ Rev.\  D {\bf 66}, 075003 (2002).

\bibitem{CP}  H. S. Goh et al.\cite{goh}; K.~S.~Babu and C.~Macesanu,
hep-ph/0505200;  S.~Bertolini, T.~Schwetz and M.~Malinsky,
  Phys.\ Rev.\ D {\bf 73}, 115012 (2006); B. Dutta, Y. Mimura and R. N.
Mohapatra, Phys.\ Lett.\ B {\bf 603}, 35 (2004);  Phys.\ Rev.\ Lett.\
{\bf 94}, 091804 (2005); W.~Grimus and H.~Kuhbock,
  arXiv:hep-ph/0607197; arXiv:hep-ph/0612132;
  C.~S.~Aulakh, arXiv:hep-ph/0607252.


\bibitem{pdecay} H.~S.~Goh, R.~N.~Mohapatra, S.~Nasri and S.~P.~Ng,
  Phys.\ Lett.\  B {\bf 587}, 105 (2004);  T.~Fukuyama, A.~Ilakovac,
T.~Kikuchi, S.~Meljanac and N.~Okada, Eur.\ Phys.\ J.\  C {\bf 42}, 191
(2005).


\bibitem{sym}  C.~S.~Aulakh, B.~Bajc, A.~Melfo, G.~Senjanovic and
F.~Vissani,  Phys.\ Lett.\ B {\bf 588}, 196 (2004);  B.~Bajc, A.~Melfo,
G.~Senjanovic and F.~Vissani,  Phys.\ Rev.\ D {\bf 70}, 035007 (2004);
T.~Fukuyama, A.~Ilakovac, T.~Kikuchi, S.~Meljanac and N.~Okada,
  Phys.\ Rev.\ D {\bf 72}, 051701 (2005).


\bibitem{gmn} H.~S.~Goh, R.~N.~Mohapatra and S.~Nasri,
  Phys.\ Rev.\  D {\bf 70}, 075022 (2004).


\bibitem{kita} See for example,
  R.~Kitano and T.~j.~Li, Phys.\ Rev.\ D {\bf 67}, 116004 (2003)
  [arXiv:hep-ph/0302073].


\bibitem{fuku} T.~Fukuyama, T.~Kikuchi and N.~Okada,
 arXiv:hep-ph/0702048, to be published in Phys.\ Rev.\ D.


\bibitem{RS}
  L.~Randall and R.~Sundrum,
  %``A large mass hierarchy from a small extra dimension,''
  Phys.\ Rev.\ Lett.\  {\bf 83}, 3370 (1999)
  [arXiv:hep-ph/9905221].


\bibitem{SUSYL} M.~A.~Luty and R.~Sundrum,
  Phys.\ Rev.\ D {\bf 64}, 065012 (2001);
  J.~Bagger, D.~Nemeschansky and R.~J.~Zhang,
  JHEP {\bf 0108}, 057 (2001);
 D.~Marti and A.~Pomarol,  Phys.\ Rev.\ D {\bf 64}, 105025 (2001).


\bibitem{dutta}   B.~Dutta, Y.~Mimura and R.~N.~Mohapatra,
  %``Neutrino masses and mixings in a predictive SO(10) model with CKM CP
  %violation,''
  Phys.\ Lett.\ B {\bf 603}, 35 (2004); Phys.\ Rev.\  D {\bf 72}, 075009
(2005).


\bibitem{D-term}
  R.~Barbieri, R.~Contino, P.~Creminelli, R.~Rattazzi and C.~A.~Scrucca,
  %``Anomalies, Fayet-Iliopoulos terms and the consistency of orbifold field
  %theories,''
  Phys.\ Rev.\  D {\bf 66}, 024025 (2002);
%
  S.~Groot Nibbelink, H.~P.~Nilles and M.~Olechowski,
  %``Spontaneous localization of bulk matter fields,''
  Phys.\ Lett.\  B {\bf 536}, 270 (2002);
  Nucl.\ Phys.\  B {\bf 640}, 171 (2002);
%
  H.~Abe, T.~Higaki and T.~Kobayashi,
  %``Wave-function profile and SUSY breaking in 5D model with Fayet-Iliopoulos
  %term,''
  Prog.\ Theor.\ Phys.\  {\bf 109}, 809 (2003).


\end{thebibliography}
\end{document}